\documentclass[preprint]{aastex61}

\usepackage{natbib}

\newcommand{\asec}{$^{\prime\prime}\!$ }

\newcommand{\kms}{ km s\ensuremath{^{-1}}}
\newcommand{\msol}{ M$_{\odot}$}

\newcommand{\htwo}{HII}

\newcommand{\netwo}{[Ne{\scshape{ii}}]~}
\newcommand{\fethree}{[Fe{\scshape{iii}}]~}

\def\degr{\hbox{$^\circ$}}
\def\arcmin{\hbox{$^\prime$}}
\def\arcsec{\hbox{$^{\prime\prime}\!$}}
\def\fdg{\hbox{$.\!\!^\circ$}}

\def\farcs{\hbox{$.\!\!^{\prime\prime}$}}

\newcommand{\radec}[4]{\ensuremath{\alpha,~\delta~(J2000)~=~17^h45^m{#1}.\!\!^s{#2},~-29\degr00\arcmin{#3}\farcs{#4}}}

\shorttitle{Extreme Blue-Shifted Ionized Gas Near Sgr A*}
\shortauthors{Royster et al.}

\begin{document}

\title{ALMA Detection of Extreme Blue-Shifted Ionized Gas Within \\
0.2 pc of Sgr A* from --480\kms~to --300\kms}

\correspondingauthor{Marc J. Royster}
\email{mjroyster@u.northwestern.edu}

\author[0000-0000-0000-0000]{M. J. Royster}
\affil{CIERA and the Department of Physics \& Astronomy, Northwestern University,
2145 Sheridan Road,  Evanston, IL 60208, USA}

\author{F. Yusef-Zadeh}
\affil{CIERA and the Department of Physics \& Astronomy, Northwestern University,
2145 Sheridan Road,  Evanston, IL 60208, USA}

\author{M. Wardle}
\affil{Department of Physics and Astronomy, and  Centre for Astronomy,
Astrophysics and Astrophotonics, Macquarie University, Sydney, NSW 2109, Australia}

\author{D. Kunneriath}
\affil{National Radio Astronomy Observatory, Charlottesville, VA, 22903, USA}

\author{W. Cotton}
\affil{National Radio Astronomy Observatory, Charlottesville, VA, 22903, USA}

\author{D. A. Roberts}
\affil{Fort Worth Museum of Science and History, Fort Worth, TX, 76107, USA}

\begin{abstract}
We have used the capabilities of ALMA to probe the ionized gas towards the Galactic center with the H30$\alpha$ mm hydrogen recombination line within 30\asec of the Galaxy's dynamical center.  The observations are made with spatial and spectral resolutions of 0\farcs46 $\times$ 0\farcs35 and 3\kms, respectively. Multiple compact and extended sources are detected in the mini-cavity region with extreme negative radial velocities ranging from --480\kms~to --300\kms~$2-3''$ ($0.08-0.12$pc) from Sgr A*.  This is the highest radial velocity of ionized gas detected beyond the inner 1$''$ of Sgr A*.  This new component is kinematically isolated from the orbiting ionized gas.  The detected gas has a velocity gradient ranging from --50 to --200\kms~arcsecond$^{-1}$~located to the southwest of Sgr A* at a position angle of $\sim$--160$^\circ$.  Previous proper motion measurements in the immediate vicinity of these extreme high-velocity ionized components have been measured and show transverse velocities that range from 313 - 865\kms.  If we assume that they are associated with each other, then the total velocity implies these components are gravitationally unbound.  In addition, we report the kinematics of cometary radio and infrared sources. These sources are diagonally distributed with a position angle of $\sim\rm 50^\circ$ within 14\asec~of Sgr A*.  We interpret the highly blue-shifted features to the SW where the mini-cavity is located and red-shifted cometary sources to the NE in terms of the interaction of a collimated outflow with an opening angle of $\sim$30$^\circ$.  An expected mass outflow rate of 2$\times10^{-7}$ or 4$\times10^{-5}$ \msol\, yr$^{-1}$ is estimated by a relativistic jet-driven outflow or collimated stellar winds, respectively.
\end{abstract}

%% See the online documentation for the full list of available subject


\section{Introduction}\label{sec:intro}
The central parsec of the Galactic center (GC) provides a unique opportunity to gain insight into the extreme physical conditions of galactic nuclei.  At the dynamical center, a supermassive black hole of 4$\times$10$^6$\msol~\citep{ghez05,gillessen09,reid04}~is coincident with the compact radio source Sgr A*. On a scale of a few parsecs (1 pc corresponds to 25$''$ at the 8.5 kpc GC distance), Sgr A* is engulfed by ionized material tracing a mini-spiral structure as well as a ring of molecular gas with a radius of 2-5 pc orbiting Sgr A* at a velocity of $\sim$100\kms~\citep{guesten87, jackson93, marshall95, latvakoski99, bradford05, herrnstein05, christopher05, ekers83, lo83}.  It is well known that the bolometric luminosity of Sgr A* due to synchrotron thermal emission from hot electrons in the magnetized accretion flow is several orders of magnitude lower than that expected from the accretion of stellar winds and ionized gas. Over the years, there have been a number of studies to address this puzzling issue using radiatively inefficient accretion flow (RIAF) models in which a fraction of the initially infalling  material accretes onto Sgr A* and the rest is driven off as an outflow from Sgr A* \citep[e.g.][]{yuan04,quataert04,shcherbakov10,wang13}.  Another class of models explaining the low luminosity from accretion onto Sgr A* (in comparison with its expected Bondi-Hoyle accretion rate from nearby gas) considers that most of the gas approaching Sgr A* is pushed away as part of a jet or outflow \citep[e.g.][]{falcke00,das09,becker11,zadeh16}. Identification of high velocity, potentially unbound ionized gas clouds tracing the interaction sites of a jet with the surrounding material would provide support for this picture. Our approach is to search for such interaction sites with millimeter (mm) hydrogen recombination line (RL) emission from ionized clouds near Sgr A*. The high sensitivity, long time baseline and broad velocity coverage observations as well as the high line-to-continuum ratio of $\sim$2 for H30$\alpha$ RL \citep{wilson12,zhao10}, provides an unprecedented opportunity to identify and probe interaction sites with extreme kinematics within a pc of Sgr A*.

One of the earliest high resolution radio studies of ionized gas in the central parsec was carried out with the VLA at 3 cm \citep{roberts93, roberts96}.  Radial velocities extending up to --280\kms\, were detected in the mini-cavity, a circular-shaped structure with a diameter of $\sim$2'' centered $\sim$3'' SW of Sgr A* \citep{zadeh90}. Similarly, infrared observations using Br$\gamma$ and \netwo lines detected a --240\kms~velocity component in the mini-cavity region \citep{lutz93,herbst93,lacy91,krabbe95}.  More recently, two studies at 1.3 cm and 1.3 mm \citep{zhao09, zhao10} used the VLA and SMA to fully map the kinematics of the mini-spiral at radio and mm, respectively.  Within a few arcseconds of Sgr A*, they reported enhanced electron temperatures approaching $\sim$1.5$\times10^4$ K. However, their studies were limited by relatively low angular resolution (1\farcs9 $\times$ 1\farcs5) and focused on radial velocities within $\pm300$\kms~which showed similar kinematics to those found in previous radio recombination line (RRL) studies. Measurements in the near-IR with the VLT detected HeI, Br$\gamma$, and \fethree lines, 2.2$''$ south of Sgr A* \citep{steiner13}. This localized ionized filament at the eastern edge of the minicavity is highly blue-shifted with a peak radial velocity $\rm v_r\sim-267$ \kms\, and velocity gradient of $\sim$200 \kms\, arcsecond$^{-1}$. There is no obvious stellar source with the same radial velocity to explain the origin of the filament. Steiner et al. (2013)  argue that the high velocity gas is shock heated and may result from the collision of the northern and eastern arms.

Here we report new ALMA observations showing multiple highly blue-shifted velocity components in the mini-cavity, one of which has a peak velocity $\rm v_r\sim-455$ \kms\, extending to $\rm v_r\sim-$480 \kms.  This is the highest blue-shifted velocity ionized cloud that has been detected, thus far, near Sgr A*. 
Combined with proper motion measurements, the new components are possibly gravitationally unbound.  We also detect RL emission from cometary sources which appear elongated in the direction of Sgr A*.  These cometary features are a subset of a number of partially resolved sources (X3, X7, F1, F2, F3, P1, P4 and the Sgr A East tower) that have been detected with radio and 3.8 $\mu$m observations \citep{zadeh12, zadeh16, muzic07, muzic10}. These have a mixture of hot dust and ionized gas and are elongated with similar position angles, $\sim60^\circ$ and $\sim-120^\circ$ \citep{muzic07, muzic10, zadeh16}.  The radial velocity and position velocity diagrams are presented. The highly blue-shifted features in the mini-cavity and the cometary sources appear to trace sites of an interaction from a collimated jet- or wind-driven outflow from the direction of Sgr A*.

\section{Observations and Data Reduction} \label{sec:obs}

The observations were carried out with the Atacama Large Millimeter/submillimeter Array (ALMA) originally as part of a multi-wavelength monitoring campaign of Sgr A*.  The calibrated Band 6 archival data (project code 2015.A.00021.S) was observed in Cycle 3 on two epochs, 2016 July 12/13 and July 18/19.  We utilized the calibrated dataset to extract the H30$\alpha$ RL in the inner parsec ($\sim$26\arcsec) of the Galactic center (GC) at 231.901 GHz with CASA.  The dataset was first combined along all spectral windows prior to deriving and applying phase self-calibration solutions three times for subsequently shorter solution intervals.  A final simultaneous phase and amplitude self-calibration solution was derived for each integration time interval before imaging the continuum.

The H30$\alpha$ data cube was constructed by first subtracting a linear continuum in the $uv$ plane and then applying the same phase and amplitude solutions derived from the continuum.  A broad velocity range of --1200\kms~to~+600\kms~was imaged, allowing a search for anomalous high-velocity ($|\rm v_r |>400$\kms) emission along with a spectral resolution of 3\kms.  A Briggs parameter of 0.5 was used to weigh the $uv$ visibilities prior to gridding which allowed a compromise between natural and uniform weighting schemes that favor extended and compact features respectively. This achieved a synthesized beam of 0\farcs46 $\times$ 0\farcs35 (PA = --67\fdg8) and full spectral resolution of 0.63\kms.  A primary beam correction was applied with a cutoff at 20\% (37\farcs2 field of view (FOV)).  The July 12/13 H30$\alpha$ image cube has $\sigma=0.4$ mJy per channel, averaged to 3\kms.

The continuum was imaged in a similar fashion but was confusion and dynamic range limited with a sensitivity of $\sigma \sim 0.8$ mJy beam$^{-1}$~and $\sigma \sim 1.6$ mJy beam$^{-1}$~for the July 12/13 and July 18/19 datasets, respectively.  Due to the relatively poor data quality of the second epoch, we chose to use the first (July 12/13) dataset only in the analysis presented here.  However, the high-velocity features described here are also detected in the July 18/19 dataset.

The peak continuum position of Sgr A* was found at \radec{40}{034}{28}{190}~in the J2000 reference frame.  In order to properly compare to other datasets, we applied a shift to the coordinate system by $+$61 and --123 mas in right ascension and declination respectively such that the position of Sgr A* could be found at \radec{40}{038}{28}{069}.  Unless otherwise noted, the positions of sources are reported as offsets with respect to this corrected position.  All images are displayed in the J2000 coordinate reference system.

\section{Results}

\begin{figure}
\centering
\includegraphics[scale=0.9]{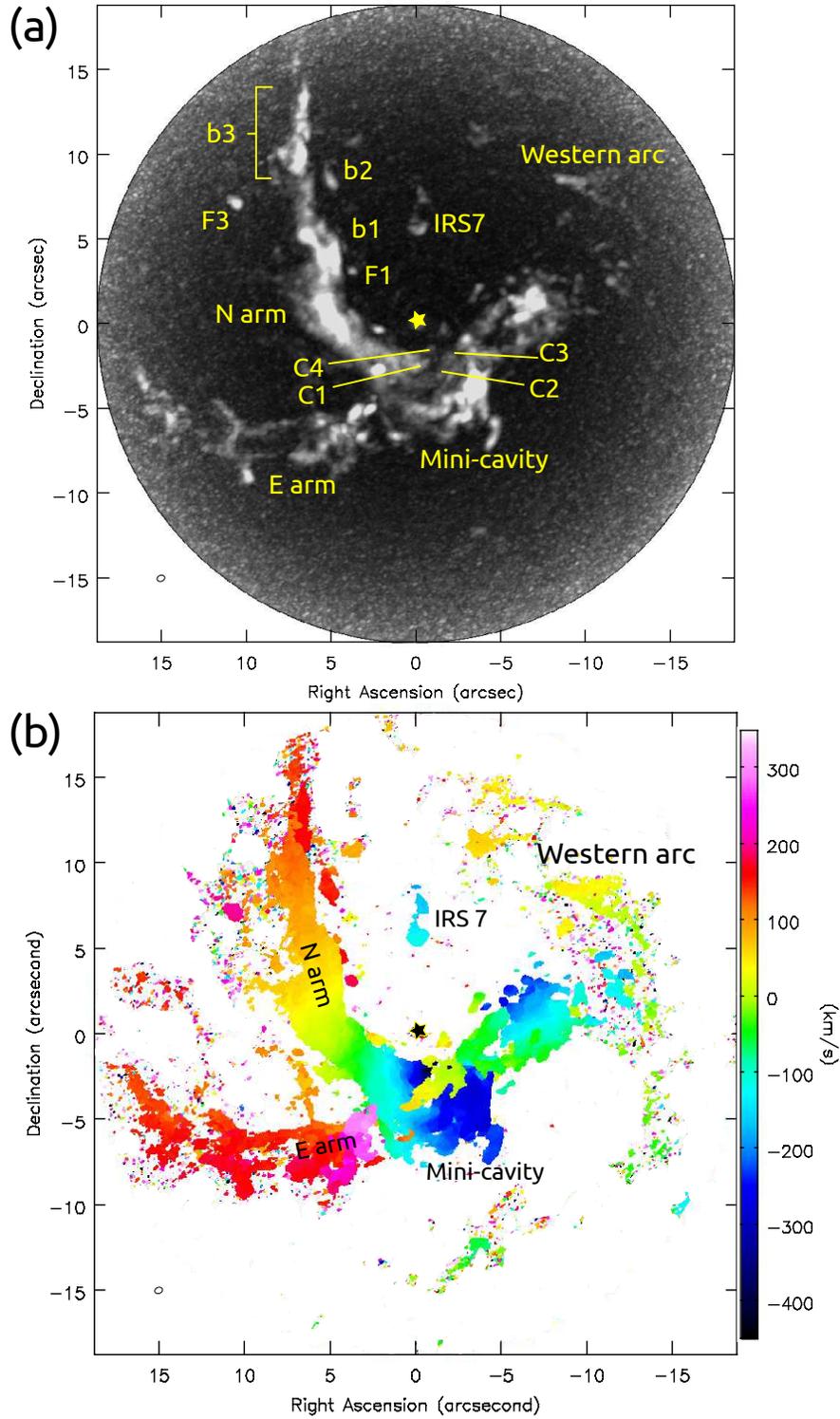}
\caption{\emph{(a)} The 231.9 GHz H30$\alpha$ peak intensity of the mapped region with synthesized beam, 0\farcs47$\times$0\farcs36 ($\rm PA = -67\fdg82$).  Coordinates are offset from the position of Sgr A* (yellow star).  
\emph{(b)} The velocity coordinate of the peak intensity for H30$\alpha$.  The position of Sgr A* is designated with a black star.}
\label{fig:map}
\end{figure}

Figure \ref{fig:map}a displays a peak intensity image of H30$\alpha$ at 231.9 GHz. Three prominent components of the mini-spiral are labeled as the northern arm, eastern arm and the western arc at the edge of the FOV. A  number of compact sources are detected, the most prominent of which is IRS 7, an M2 supergiant star \citep{wollman82, lebofsky82}, with a head-tail structure \citep{zadeh91,serabyn91} at PA $\sim$ 0$^\circ$~from Sgr A*. Figure \ref{fig:map}b shows the velocity of the peak emission of the H30$\alpha$ recombination line from the mini-spiral on the same scale and resolution as Figure \ref{fig:map}a. The kinematics of the mini-spiral traced in Figure \ref{fig:map}b are very similar to past radio and infrared measurements \citep[e.g.][]{roberts93,roberts96,zhao09,zhao10,irons12}. Broadly speaking the N arm shows red-shifted radial velocity components $\rm v_r\sim$100\kms~which gently decreases to $\rm v_r\sim0$\kms~from north to south and then makes a large velocity jump to $\sim$-300\kms, as the gas approaches Sgr A* from the east. The velocity gradient of the eastern arm has an opposite trend, becoming increasingly red-shifted before it changes direction and becomes blue-shifted ($\sim$ $-200$\kms) near Sgr A*.  

The fourth prominent component of the mini-spiral is within the inner few arcseconds of Sgr A*, the so-called ionized bar.  The ionized bar is in fact the high negative velocity component of $-200$\kms~which lies closest to Sgr A*. There is a ridge of highly blue-shifted gas, with a U-shaped structure (i.e., the mini-cavity) depressing the continuum emission associated with the ionized bar, where the northern and eastern arms appear to converge. Although the dominant emission is blue-shifted, there is also zero\kms\, gas that is detected throughout this region. Our observations show for the first time highly blue-shifted ionized material near Sgr A*, as described below. 

\subsection{Mini-cavity}
\begin{figure} 
\includegraphics[width=\textwidth]{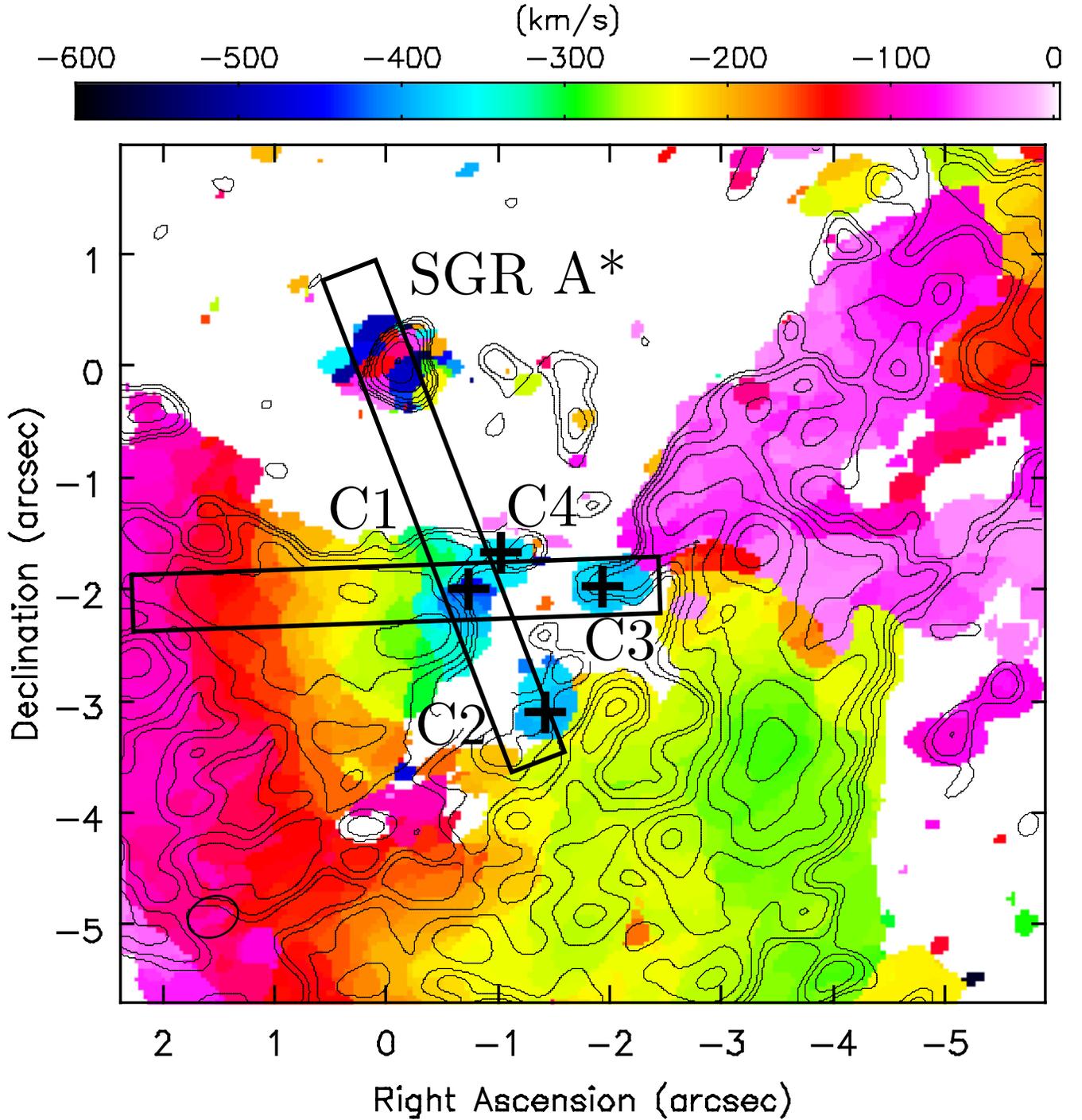}
\caption{
H30$\alpha$ velocity coordinate for the peak intensity ($> 0.2$ Jy beam$^{-1}$) of blue-shifted emission in the mini-cavity region.  The contours are integrated intensity beginning at 0.1 Jy beam$^{-1}$~and increasing by factors of 1.5.  The blue/teal colored emission corresponds to velocities less than roughly --350\kms.  The four `+' signs are the positions of C1-C4.  Each of the two rectangular boxes represent a position-velocity slice found in Figure \ref{fig:mcpv}~with a common offset position of \radec{39}{98}{30}{25}.  
}
\label{fig:mcmap}
\end{figure}

We have found four highly blue-shifted velocity components, C1 to C4, along the western and eastern walls of the mini-cavity. The kinematics of these sources exceed those of the orbiting gas of the ionized bar by $\sim$200\kms. Figure \ref{fig:mcmap}~shows the distribution of the peak velocity of the mini-cavity region in H30$\alpha$.  The four aforementioned sources labeled and distinguishable by their highly blue-shifted $\sim$--400\kms~emission (blue/teal in the image). The emission from the orbiting gas associated with the ionized bar is shown in red and pink. 

\begin{figure} 
\includegraphics[width=\textwidth]{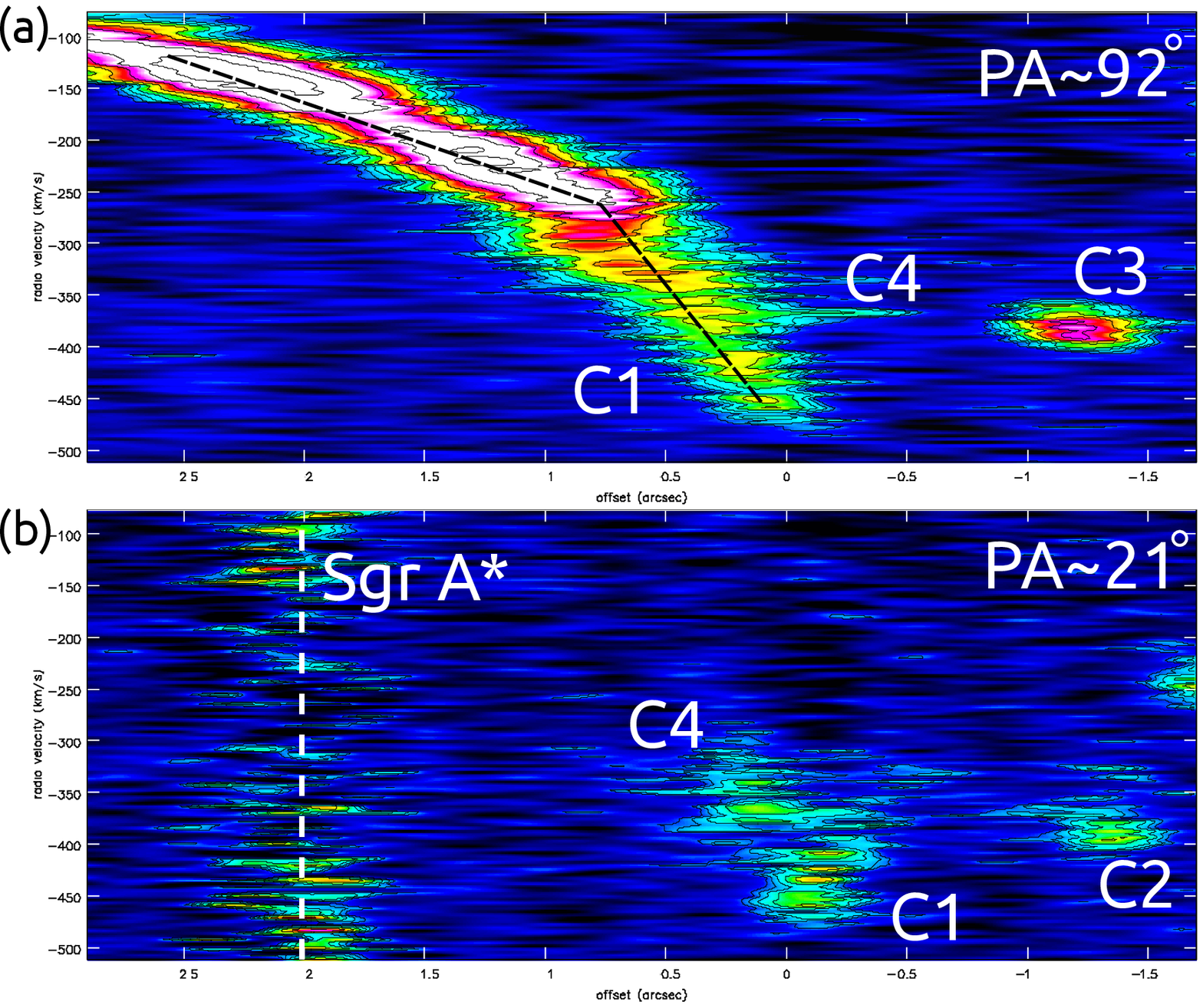}
\caption{Position-velocity slices with zero offset position taken as \radec{39}{98}{30}{25}, while the length of each slice is 4\farcs7 and integrated width is 0\farcs44.  The given position angle (in degrees) is printed in the top right corner of each and corresponds to a box in Figure \ref{fig:mcmap}.  The position of Sgr A* is represented by a white vertical dashed line in \emph{(b)}.  Contours begin at 1 mJy beam$^{-1}$~and increase by factors of 1.3.}
\label{fig:mcpv}
\end{figure}

\begin{figure} 
\includegraphics[width=\textwidth]{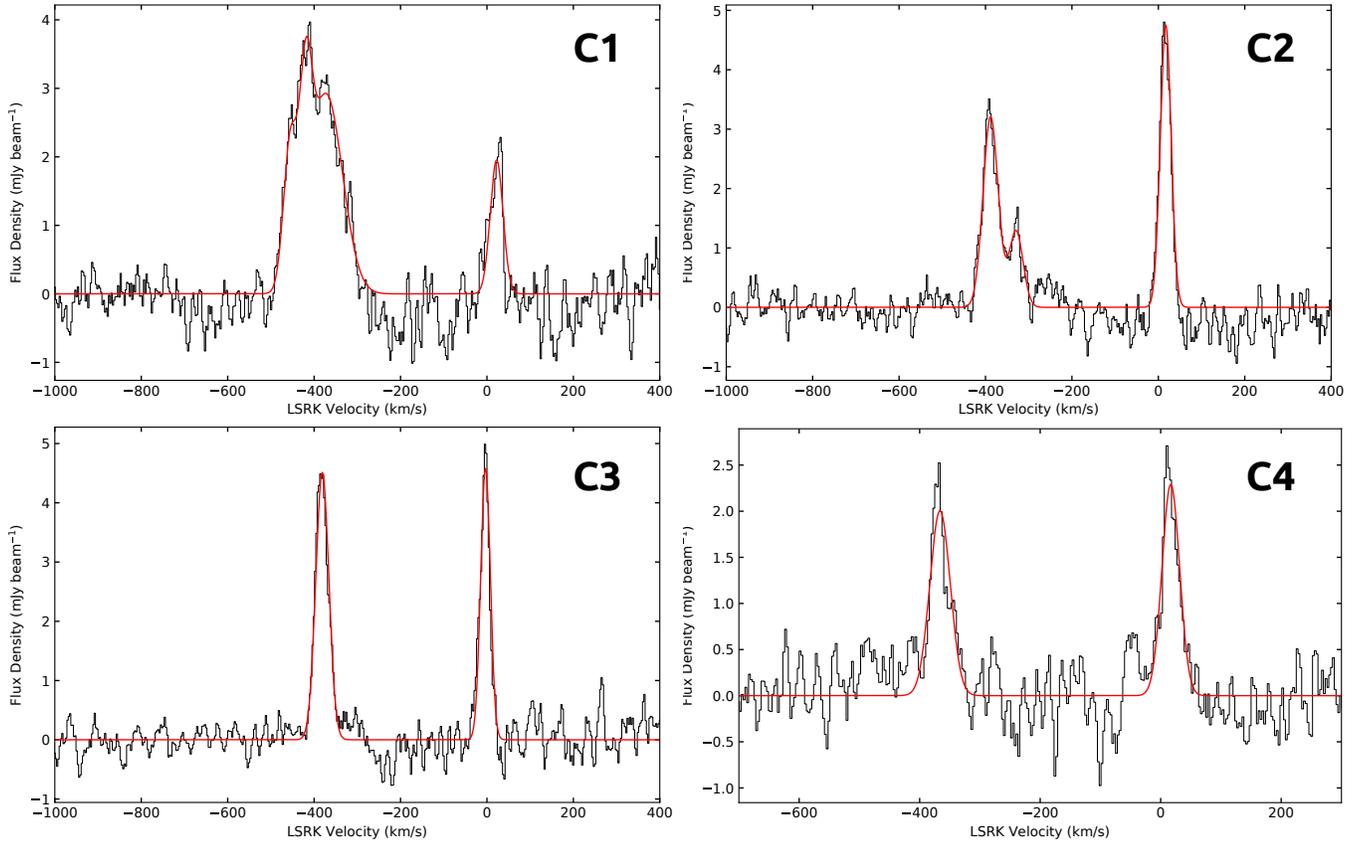}
\caption{ 
H30$\alpha$ spectra for each of the four blue-shifted sources discussed.  Each spectra is integrated over one synthesized beam (0\farcs46 $\times$ 0\farcs35) with spectral resolution of 3\kms.  The fitted parameters of the blue-shifted components are given in Table \ref{tab:table}.  The $\sim$ 0\kms~component is discussed in the text where applicable.
}
\label{fig:cspec}
\end{figure}

The position-velocity (PV) diagrams along two slices (arrows in Figure \ref{fig:mcmap}) that pass though C1 to C4 are shown in Figure \ref{fig:mcpv}. Figure \ref{fig:mcpv}a (corresponding to the nearly horizontal arrow in Figure \ref{fig:mcmap}) shows the kinematic separation of C1, C4 and C3 from the orbiting gas with lower negative velocities. Figure \ref{fig:mcpv}b shows the PV diagram of C1, C2 and C4 indicating that these sources are also kinematically isolated despite having similar~$\rm v_r <-350$\kms.  The individual velocity profiles of each of the four components are shown in four panels in Figure \ref{fig:cspec}. C1 shows the highest negative velocity, consisting of (at least) three blue-shifted components.

\begin{figure} 
\includegraphics[width=\textwidth]{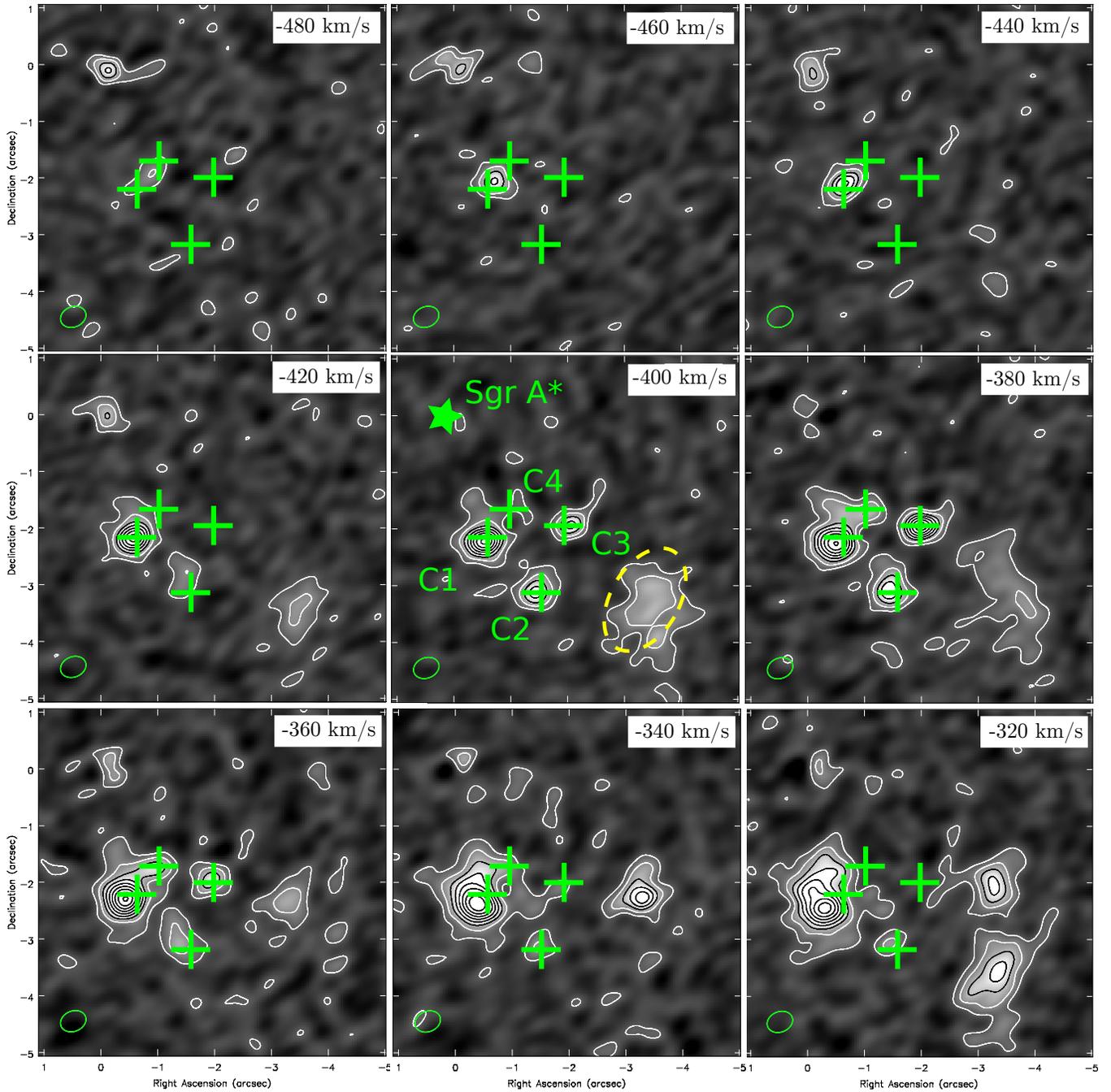}
\caption{ Channel maps of H30$\alpha$ line emission from the mini-cavity (6\asec $\times$ 6\asec) showing the most negative velocity components with velocity intervals of 20 \kms.  Contours begin at 1.9 mJy beam$^{-1}$~and increase by factors of 1.2.  The `+' signs are the positions of C1-C4 and the star is position of the Sgr A*.  The dashed ellipse in the middle panel is the H30$\alpha$~emission mentioned in Section \ref{sec:mc}.}
\label{fig:mcpost}
\end{figure}

Figure \ref{fig:mcpost} shows nine channel maps of H30$\alpha$ emission between --380 and --480\kms.  These maps show clearly the blue-shifted emission is extended but is confined to the mini-cavity region.  In addition to the aforementioned C sources an additional ridge (yellow dashed ellipse) of $\sim$ --400\kms~emission is seen in the SW corner of the mini-cavity.  Because it is coincident with the orbiting gas (seen at velocities $\lessapprox$ 320\kms) it is not visible in Figure \ref{fig:map}b.  The ridge of emission is most likely He30$\alpha$ as it is separated by the expected value of 122\kms~to H30$\alpha$~emission at --272\kms.  The He30$\alpha$~emission was also detected by \cite{tsuboi17b}.  We find $I_{\rm He30\alpha}/I_{\rm H30\alpha}\sim0.8$ in this region which is typical for \htwo~regions~\citep{wilson12}.

The positions and physical properties of the individual sources C1 - C4 are described below and are summarized in Table~\ref{tab:table}.  The first column of Table \ref{tab:table}~is the given source name as described throughout the text.  Column 2 is the offsets from the position of Sgr A* (see Section \ref{sec:obs}) in right ascension and declination respectively.  Columns 3 and 4 are the projected offset and position angle with respect to Sgr A*.  Column 5 is an upper limit to the geometric mean of the source.  Columns 6, 7 and 8 are the peak intensity, central velocity and the FWHM of the spectra integrated over one synthesized beam respectively.  Note that only relevant (in this case `high-velocity') components are listed for each source.  Column 9 is the lower-limit of the electron density for assumed electron temperatures of $7.5 - 12.5\times10^3$ K (see Section \ref{sec:density}).  For the head-tail sources (see Section \ref{sec:comet}), column 10 lists the angular length of the structure and column 11 is the difference in velocity between the head and tail.

\subsubsection{C1 \& C4} \label{sec:c1c4}
The closest sources to Sgr A* with the highest blue-shifted velocities are C1 and C4 which lie at the northeastern corner of the mini-cavity. C1 is the most prominent of the four high velocity components defined by the most extreme blue-shifted emission. An elliptical Gaussian was fit to the RL integrated intensity from --460 - --450\kms.  The result was a peak found at $\Delta\alpha=-0\farcs68$, $\Delta\delta=-2\farcs06$ with a projected distance and PA from Sgr A* of 2\farcs17 ($\sim$0.09 pc) and --163$^\circ$ respectively.  The top-left panel of Figure \ref{fig:cspec} shows the spectrum at this position. Three velocity components characterize the high velocity emission, one of which has the highest negative velocity (--455\kms) and a line width (FWHM) $\Delta \rm v_{\mathrm{r}} = 37$\kms, as listed in Table \ref{tab:table}.  This is the highest velocity gas found towards the inner parsecs of the GC.  In addition the source is notable for a breadth of emission that extends from approximately --460\kms~to --300\kms~in a $\sim$1\farcs2 square region.

The position of C1 is still $\sim$ 0\farcs5 NE of the highest velocity emission found by \cite{steiner13} in the IR.  At their 
position, we find emission that ranges from roughly --415 to --240\kms~compared to their reported detection at 
--378\kms~with \fethree.  In addition, \cite{tsuboi17a}~detected H42$\alpha$~emission at --380\kms~SW of this position (2\farcs8 from Sgr A*) and attribute it as the counterpart to the high velocity Br$\gamma$ emission detected by \cite{steiner13}.

C4 is defined by the integrated emission from --370 - --360\kms~and is found to be projected about 0\farcs5 NE of C1 and is the closest high velocity gas to Sgr A* at a distance of 2\farcs0 ($\sim$0.08 
pc) from Sgr A*.  The emission peaks at --366\kms~which is less than all three of the fitted components of C1 but is 
most likely associated with the broad --372\kms~C1 gas.  Both C1 and C4 have a low velocity component near 20\kms~with 
$\Delta \mathrm{v}_\mathrm{r}\sim30$\kms.

Figure \ref{fig:mcpv}b shows a PV diagram that connects the position of Sgr A* with C1.  Because of the non-thermal contribution from Sgr A*, the continuum is poorly fitted at this position.  As a result, there appears to be emission, in every channel (the vertical dashed white line).  The emission due to C1 is apparent in the panel (located at 0\asec~offset) as is that of C4, however there is not a significant contribution of ionized gas between the projected positions of C1/C4 and Sgr A*.  

A summary of the velocity gradient is best seen in the PV diagram in Figure~\ref{fig:mcpv}a.  The velocity gradient in the panel extends from the aforementioned C1 contribution of --460\kms~to --100\kms~from the N arm over a projected distance of $\sim$3\asec.  This yields an overall change of 120\kms~arcsec$^{-1}$ or 5.0\kms~pc$^{-1}$.  However, a noticeable kink in the ionized emission appears in the PV diagram at an offset of roughly 0\farcs9 (to the east).  If only positions to the east of C1 (further along the Northern Arm) are considered, the observed gradient is reduced to 71\kms~arcsec$^{-1}$~ which is more consistent with the orbiting gas of the mini-spiral.  However, within 0\farcs9 of C1, the velocity gradient is measured as 233\kms~arcsec$^{-1}$~or roughly 10\kms~pc$^{-1}$.  C4 is distinguishable as a small horizontal extension from this increase in velocity gradient

\subsubsection{C2} \label{sec:c2}

The C2 source is found towards the projected interior of the mini-cavity just north of the southern ridge at $\Delta\alpha=-1\farcs44$, $\Delta\delta=-3\farcs11$ with a projected angular distance from Sgr A* of 3\farcs4.  The position was determined with RL integrated intensity from --400 - --390\kms.  Notably, a PA from Sgr A* of roughly --155\degr~intersects all of C1, C2 and C4.  This is evident in the PV diagram found in the bottom panel of figure~\ref{fig:mcpv} where the ionized emission of C2 is seen at an offset of roughly 1\farcs5.  The 2 cm radio continuum source $\eta$~of \cite{zadeh90} is coincident with C2. 

The H30$\alpha$ spectra of C2 is found in Figure \ref{fig:cspec}.  Two blue-shifted components are measured with the most extreme found with a center velocity of $\rm v_r=-388$\kms, FWHM of $\Delta \rm v_\mathrm{r} = 41$\kms~and peak intensity of 3.2 mJy beam$^{-1}$.  The kinematics are remarkably different than the neighboring ionized orbiting gas found just south of C2 where velocities are greater than --250\kms.  Unlike the C1 source, there appears to be no kinematic connection between the emission and the orbiting gas seen in PV diagrams.  However, similar to C1 (and C4), a low-velocity component is found at $\sim$ 20\kms.

\subsubsection{C3} \label{sec:c3}

The third of the extreme blue-shifted sources is found at $\Delta\alpha=-1\farcs99$, $\Delta\delta=-1\farcs96$ just east of the northwestern 
corner of the mini-cavity 2\farcs8 from Sgr A*.  We have defined C3 by the RL integrated intensity from --390 to --380\kms.
Similar to C2, C3 appears to be kinematically isolated from 
the orbiting ionized gas and more compact than C1 (see Figure \ref{fig:mcpv}).  
C3 has a peak component at 
$\rm v_r=-381$\kms, with $\Delta \rm v_r=35$\kms, and intensity of 4.64 mJy beam$^{-1}$~(spectra shown in Figure \ref{fig:cspec}~and is in the vicinity of the $\zeta$ radio continuum source detected in \cite{zadeh90}).  The 
position angle with respect to Sgr A* varies from that of C1 and C2, and instead is found at roughly 
--135\degr.  This is a difference of nearly 30\degr~from the C1 PA.
Consistent with the other mini-cavity sources, a low-velocity component is detected, 
but  at --3.3\kms~instead of 20\kms.  

\subsubsection{The Gas Density} \label{sec:density}

To approximate the electron density ($n_e$) of the extreme blue-shifted emission, nominal values of 7.5$\times10^3$ K and 12.5$\times10^3$ K were used for the electron temperature.  The electron temperature is known to be elevated near the mini-cavity \citep[see][]{roberts93,zhao10}, deviating from the typical values of 7.5$\times10^3$ K for \htwo~regions.  The size of the sources were determined by fitting an elliptical Gaussian on the plane of the sky where the blue-shifted emission was detected.  When the source was resolved we extracted the source size deconvolved from the beam otherwise it was convolved with the beam.  This yields an upper-limit on the size of the source or a lower-limit on the $n_e$ as calculated below, assuming a path length equal to this upper value.

The line intensity at 231.9 GHz can be expressed as \citep{wilson12}:

\begin{equation}
\mathrm{T}_\mathrm{L} = 1.92\times10^3 \left(\frac{\mathrm{T}_\mathrm{e}}{\mathrm{K}}\right)^{-3/2} 
                   \left(\frac{\mathrm{EM}}{\mathrm{cm}^{-6}\,\mathrm{pc}}\right)^{-1}
                   \left(\frac{\Delta \nu}{\mathrm{kHz}}\right) \mathrm{K}.
\end{equation}

\noindent Solving for the emission measure (EM) while converting to velocity space and flux density,

\begin{equation} \label{eqn:em}
\mathrm{EM} = 5.71\times10^{-2} \left(\frac{\mathrm{T}_\mathrm{e}}{\mathrm{K}}\right)^{3/2} 
                     \left(\frac{\mathrm{S}_\mathrm{L}}{\mathrm{mJy}}\right)
                     \left(\frac{\Delta \rm v_r}{\mathrm{km}\, \mathrm{s}^{-1}}\right) \mathrm{cm}^{-6}\,\mathrm{pc},
\end{equation}
dividing by the path length and taking the square root then gives us the $n_e$.  Most of the RL emission observed was best characterized with multiple velocity components.  Thus to apply equation \ref{eqn:em}~we interpreted the last two factors as the sum of the product of each (blue-shifted) component, $\sum{(\mathrm{S}_\mathrm{L} \Delta \rm v_r)}$.

The estimated (lower limit) $n_e$ is given in Table \ref{tab:table}~for both values of the assumed T$_{\mathrm{e}}$~with path length taken to be identical to the source size projected on the sky.  C1 is found to have a much higher density with a value of $\sim(3.4 - 5.0) \times 10^4$ cm$^{-3}$ than the other mini-cavity sources.  C3 and C4 have a much lower range of 1.4 - 2.7~$\times 10^4$ cm$^{-3}$.  The $n_e$ of C2 is comparable to C1 but it should be noted that the assumed size of C2 is nearly half that of C1.

To approximate the \emph{total} mass of the extreme blue-shifted emission coincident with the mini-cavity, the entire region that contained significant emission from --500 to --300\kms~was integrated.  The region was equivalent to 11.7 beams ($1.5\:\rm arcsec^2$) at roughly $\Delta\alpha=-1\farcs05$, $\Delta\delta=-2\farcs36$.  This velocity range accounts for the peaks of negative velocity components as well as the width seen due to thermal and turbulent broadening.  The resultant spectra of the blue-shifted emission was best fit with two Gaussians at --353 and --264\kms.  The more negative component has a peak and FWHM of 1.8 mJy beam$^{-1}$~and 131\kms~respectively.  Applying equation \ref{eqn:em}, we determine an EM of roughly $1.3 \times 10^{7}$~cm$^{-6}$~pc from this negative component.

The geometric mean of the area of the integrated region was used as the path length, this corresponds to 1\farcs2 or 0.05 pc, which is roughly the diameter of the mini-cavity.  If a nominal value of $10^4$ K for electron the temperature is assumed, then an average electron density of 1.6$\times 10^4$ cm$^{-3}$ and total mass of 0.052\msol~for the highly disturbed ionized gas in the mini-cavity is determined.

\subsection{Cometary Sources}  \label{sec:comet}
\begin{figure}
\includegraphics[width=\textwidth]{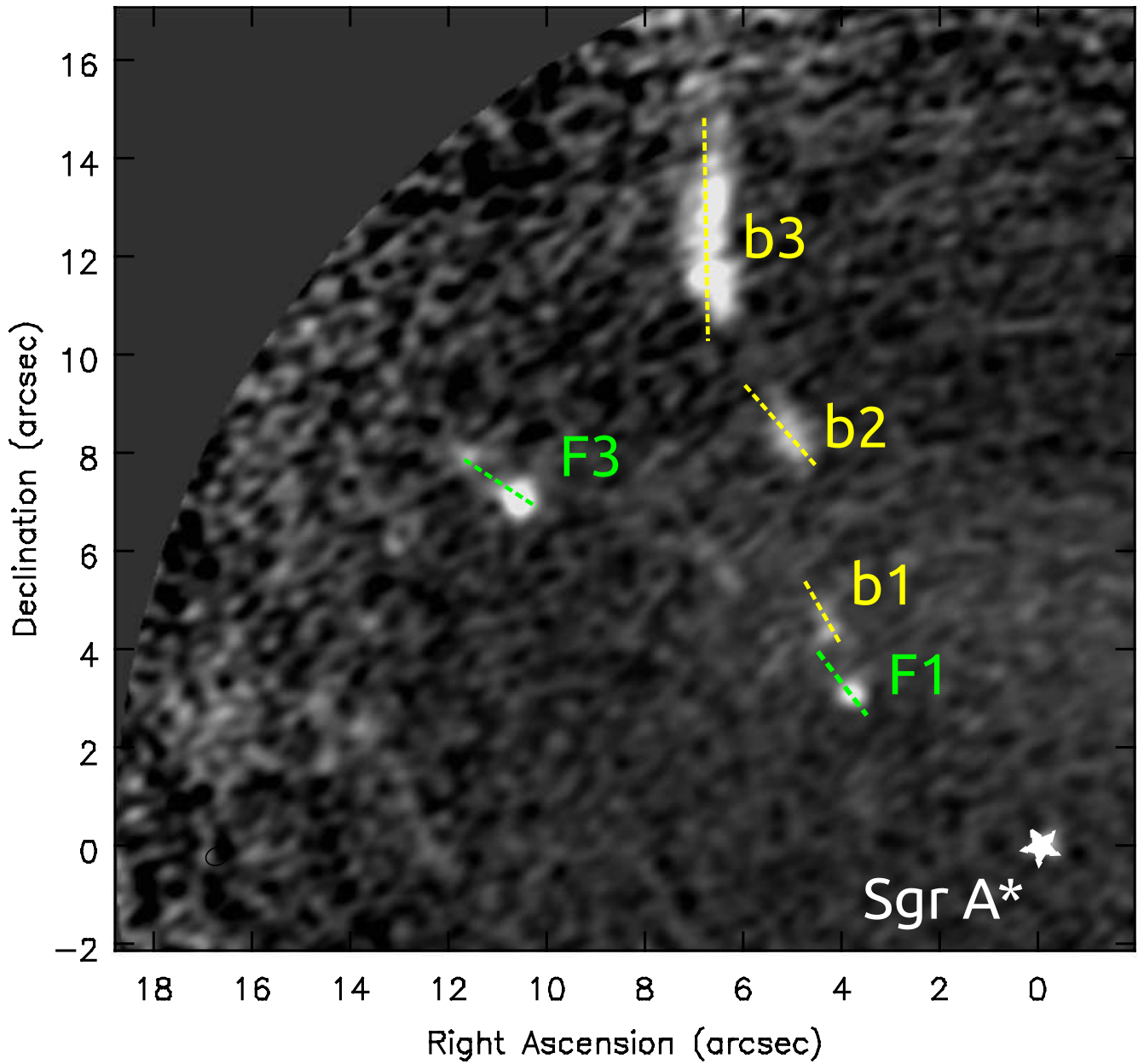}
\caption{
A H30$\alpha$ integrated intensity map from 150\kms~to 190\kms.  All the F and b sources are visible in this range.  
A dashed line represents a corresponding PV diagram in Figures \ref{fig:fpvs}~and \ref{fig:bpvs}.  The F1 and F3 PAs (lime colored) are 36\degr~and 57\degr~respectively.  The b1, b2, and b3 PAs (banana colored) are 30\degr, 41\degr, and 0\degr~respectively.
}
\label{fig:fbmap}
\end{figure}

\begin{figure}
\includegraphics[width=\textwidth]{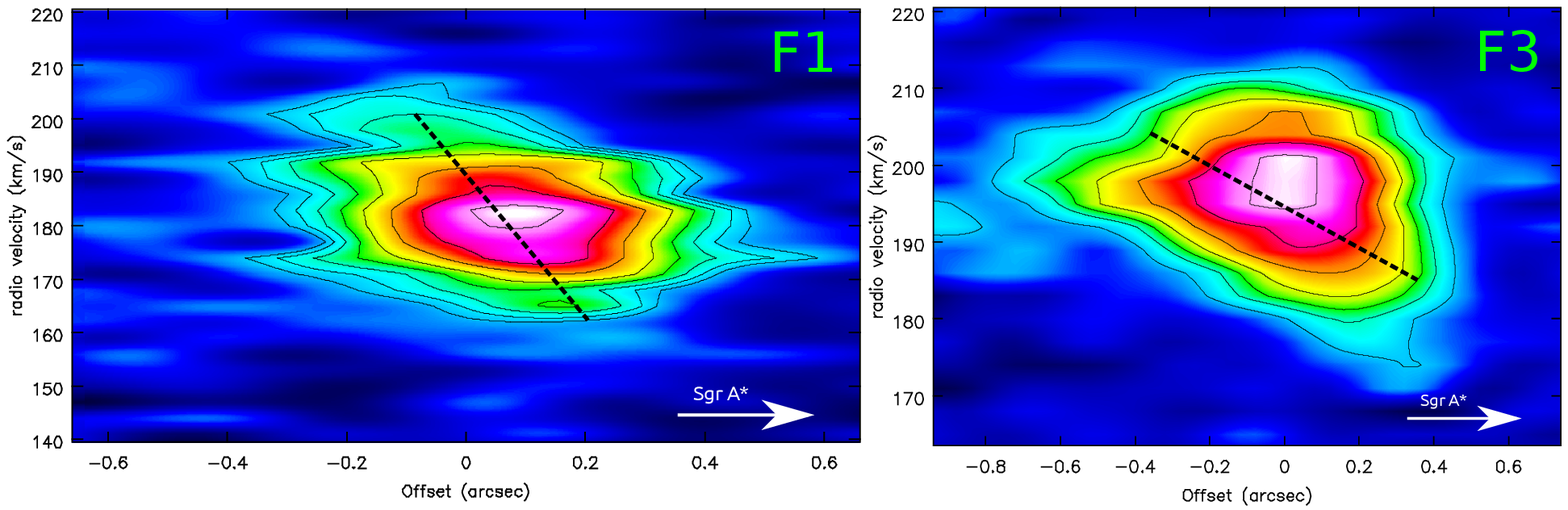}
\caption{H30$\alpha$ position-velocity (PV) diagrams for the F sources to the NW of Sgr A*.  
\emph{Left}: H30$\alpha$ F1 PV diagram.  The zero offset position (\radec{40}{33}{25}{05}) was determined by fitting the H30$\alpha$ peak position.  The position angle (PA) of the slice was taken to be that of the $L'$-band cometary source \citep{muzic10}~of 36\fdg22.  Sgr A* is at a PA of 51\fdg7 from the zero offset position (F1).  The width of the PV slice was 0\farcs44.  Contours begin at 1.2 mJy beam$^{-1}$~and increase by factors of 1.25. \emph{Right}: H30$\alpha$ F3 PV diagram with zero offset at \radec{40}{83}{21}{12}~found using integrated H30$\alpha$ intensity and with a PA with respect to Sgr A* of 56\fdg5~(also used as the PA of the slice). Contours begin at 0.7 mJy beam$^{-1}$~and increase by factors of 1.25.  A width of 0\farcs6 was used.
}
\label{fig:fpvs}
\end{figure}

\begin{figure}
\includegraphics[width=\textwidth]{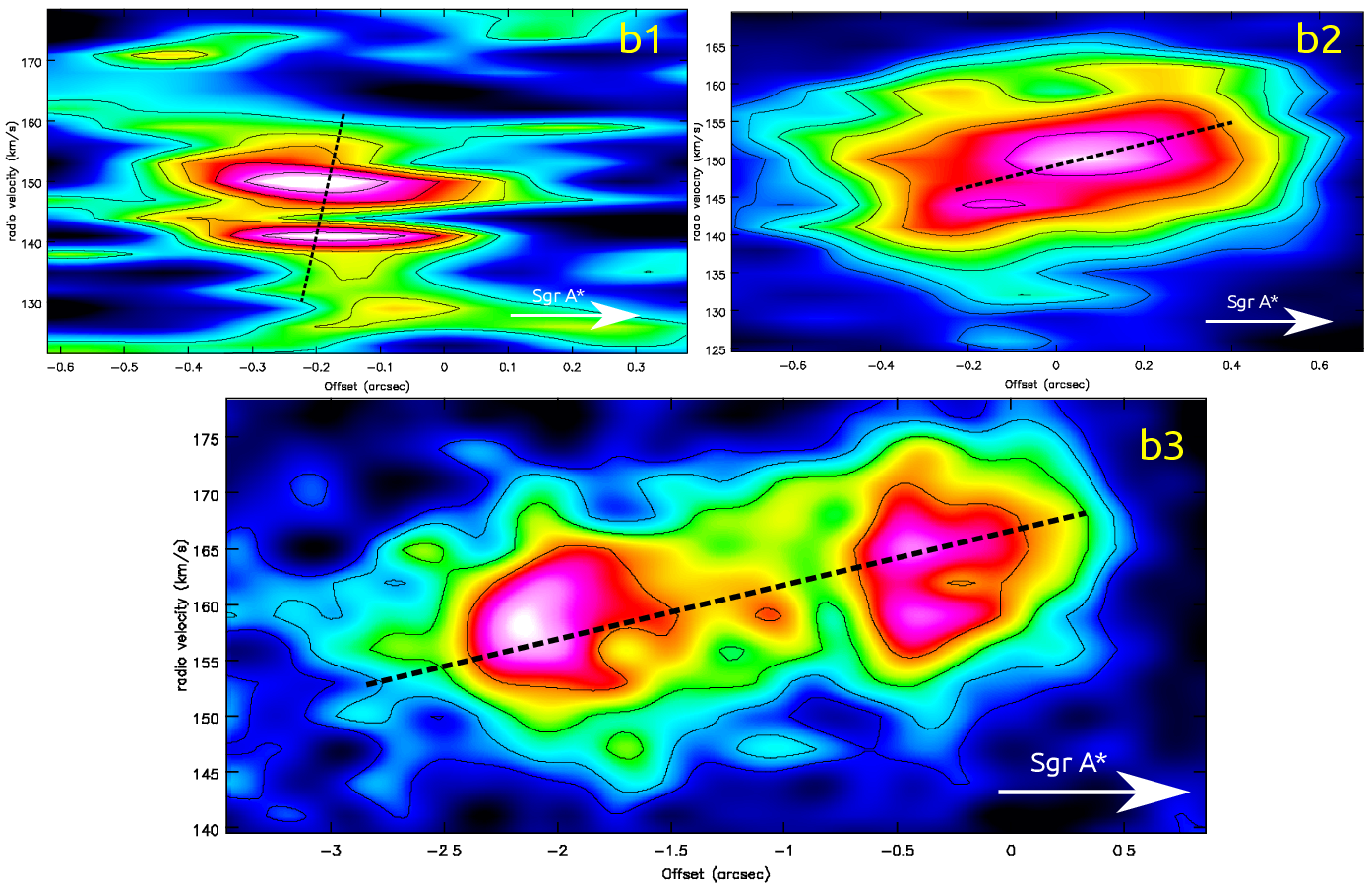}
\caption{H30$\alpha$ position-velocity (PV) diagrams for the b sources to the NW of Sgr A*. 
\emph{Top-Left}: H30$\alpha$ b1 PV diagram with zero offset at \radec{40}{36}{23}{80}~found
fitting H30$\alpha$ emission at 170\kms.  A PA=29\fdg95 matching the orientation of the cometary source.  
Contours are based on 2$\sigma =2\times 0.7$ mJy beam$^{-1}$~and increase by multiples of 1.25.  A width of 0\farcs4 was used. \emph{Top-Right}: H30$\alpha$ b2 PV diagram with zero offset coincident with the peak position of 
150\kms~emission at \radec{40}{42}{19}{80} with PA=41\fdg40. 
Contours begin at 0.8 mJy beam$^{-1}$.  
A width of 0\farcs75 was used.
\emph{Bottom}: H30$\alpha$ b3 PV diagram with zero offset at \radec{40}{56}{16}{32}.  A width of 1\farcs0~was used and the cut is along a 0\degr~PA.
}
\label{fig:bpvs}
\end{figure}

To the NE of Sgr A*, a number of previously detected cometary sources are detected in H30$\alpha$ RLs.  Figure \ref{fig:fbmap}~is an integrated intensity map from 150 to 190\kms~with the detected cometary sources labeled in relation to Sgr A*.  PV diagrams of the five sources can be found in Figures \ref{fig:fpvs}~and \ref{fig:bpvs}.

\subsubsection{F Sources}
F1 and F3 are known cometary sources pointed in the direction of Sgr A*. These cometary tails are associated with dusty stars that have been detected in the radio \citep{zadeh16}. The PA (with respect to Sgr A*) of F1 and F3 (52$^\circ$ and 55$^\circ$ respectively) is similar to two other IR-identified cometary sources, X3 and X7, found to the SW of Sgr A* \citep{muzic07, muzic10}~as well as the mini-cavity source C3 discussed in Section \ref{sec:c3}.  

Radial velocities of these dusty sources are derived from their H30$\alpha$ emission and the parameters that characterize these sources are described in Table \ref{tab:table}. Electron densities of $\sim$ $2\times10^4$ cm$^{-3}$ are determined if an electron temperature of $7.5\times10^3$ K is assumed.  The PV diagrams of F1 and F3 (Figure \ref{fig:fpvs}) show that the tail, pointed in the direction away from Sgr A*, is more red-shifted than the head of the source.  Position angles of 36\degr~and 55\degr~were adopted for the PV slices of F1 and F3 respectively which matched the morphological orientation of the cometary sources in the $L^\prime$-band data \citep{muzic07}.  For the F1 cometary source, the kinematics vary from 179\kms~at the head to 187\kms~at the tail across an angular distance of 0\farcs5.  Similarly, F3 varies from 191\kms~to 198\kms~over 0\farcs7. 

\subsubsection{b Sources}
There are also a number of additional cometary sources seen in ionized gas, b1, b2 and b3 in Figure \ref{fig:fbmap}, that run nearly parallel to the N arm of the mini-spiral.  The PV diagram of each can be found in \ref{fig:bpvs}.  It is not clear if any of the b sources have stellar counterparts.  The general velocity structure differs from that of the F sources which have a red-shifted tail.  The overall direction is pointed towards Sgr A*, however the position angle of the head-tail structure does differ from F1 and F3, as they are rotated towards the direction of F1.

The first of the sources, b1, is separated from F1 by roughly 1.3\asec~and is blue-shifted with respect to F1 with $\rm v_{\mathrm{r}}\approx165$\kms.  The source is the weakest of the discussed cometary sources with a peak intensity of 2.92 mJy.  Over an angular distance of 0\farcs3 we find a shift in velocity of 6.6\kms~with a red-shifted tail, similar to the F sources.

The remaining two, b2 and b3, both have a velocity gradient with a red-shifted head.  These appear to potentially be part of a coherent source distinct from the N arm whose kinematic structure opposes b2 and b3 (see Figure \ref{fig:map}b).  b2 is roughly 10\asec~from Sgr A* with a central velocity of 152\kms.  Elongated over an angular distance of $\sim$ 0\farcs9 and $\rm PA = 41$\degr, b2 varies by 3.7\kms.  b3 is found further to the north, 13\farcs5 from Sgr A* and extended over 2\asec~- 3\asec, coincident with the northwestern ridge of the N arm.  A velocity gradient of 3.9\kms~arcsec$^{-1}$ is measured.  

\section{Discussion} \label{sec:discussion}

\subsection{Unbound Mini-cavity} \label{sec:mc}
We have detected highly blue shifted ionized gas with radial velocities ranging between --480 and --300\kms. This gas is distributed in a region between 2\farcs0 and 3\farcs5 from Sgr A*. The high velocity features are partially resolved and are physically associated with the mini-cavity ridge with a 2\asec~circular-shaped diameter.  The mini-cavity is embedded within the ionized bar which is dominated by the orbital motion of the gas.  Proper motion measurements of the eastern and western edges of the mini-cavity show transverse velocities moving southwest and west, respectively \citep{zadeh98,zhao09}. In one study, the regions roughly coincident with the edges of the eastern and western walls of the mini-cavity show transverse velocities ranging between $-496\pm97$ and $804\pm255$\kms~\citep[Boxes 2 and 8 in Table 1 of][]{zadeh98}. 

More recent proper motion measurements by \cite{zhao09}, with much improved signal-to-noise data, showed consistently high transverse velocities in the mini-cavity region.  When combined with the high radial velocities of ionized gas detected here, the possibility of unbound components becomes likely.  The ionized gas is gravitationally responding to a $4\times10^6$ \msol~central source, Sgr A*.  Although a central stellar population could also contribute, the mass profile of the inner parsec is flat as derived from stellar kinematics \citep{fritz16}.  Thus, the escape velocity at a \emph{projected} distance of 0.09 pc (2\farcs2) is then $\sim$ 620\kms. For C1 to be unbound then, it must have a transverse velocity $\gtrsim420$\kms. Note that the actual requirement is less as the true distance from Sgr A* likely exceeds the projected distance.

Alternatively C1 may be contained in an eccentric Keplerian orbit around Sgr A*, $\rm v_{\rm orbit}=\sqrt{\frac{GM}{r}(1+e)}\sin i$.  If we make the assumption of an orbital inclination of $i = 90\degr$,~then $\rm v_{\rm orbit} = \rm v_r = 455$\kms~which requires an eccentricity of $\rm e\gtrsim 0.11$.  However it should be noted that in addition to the assumption of zero transverse velocity, the orbit would also be at perihelion in this scenario.

As an example of proper motions, consider the K25 and K33 sources \citep[see Table 2 of][]{zhao09}, which are 0\farcs52 and 0\farcs41 away from C1 (within one resolution element) with transverse velocities of 865\kms~and 313\kms~respectively.  When added to the --455\kms~radial component of C1 we find a total velocity of 988\kms~and 552\kms~for K25 and K33 respectively.  Thus, if we assume C1 has similar transverse velocities to these closest \htwo~knots studied in \cite{zhao09}~then C1 is unbound.  In the case of C1 being coincident with K25, this would be true considering the projected distance alone.

The other mini-cavity sources (C2 and C3) have similar escape velocities of 500 and 540\kms~respectively at their projected distances. C3 is found adjacent (0\farcs22 apart) to the K32 source of \cite{zhao09}~with a reported transverse velocity of 656\kms, which already places it in the unbound regime.  The K41 source of \cite{zhao09}~is nearly coincident with C2 (0\farcs13 away) and has a reported transverse velocity of 215\kms.  When added to the measured C2 radial velocity a total velocity of 444\kms~is found.  Thus, if the projected distance of C2 is the true distance, it is not necessarily unbound.  However if a true distance of at least 0.18 pc is assumed (versus the projected distance of 0.14 pc), the velocity of the ionized gas would exceed the escape velocity.  Such a true distance would require an inclination of 38$^\circ$~between Sgr A* and C2.

\subsection{Origin of High Velocity Gas}
\cite{steiner13} did not find any stellar sources with radial velocities similar to high velocity ionized gas. The large extent and multiple velocity components of blue-shifted gas are consistent with the picture that stars are not responsible for the origin of high velocity gas.  The origin of high velocity gas is argued to be due to collisions between the gas of the N and E arms \citep{steiner13,zhao09}. However, the U-shaped morphology of the mini-cavity embedded in the bar is difficult to explain in this picture both spatially and kinematically. Both the N  and E arms have velocities that are lower than than the high velocities that are reported here.  
 
The new measurements show that the ionized bar is kinematically disturbed in the region traced by the minicavity ridge. Also, the highest ratio of \fethree to radio continuum in the inner 5$''$ of Sgr A* coincides with the minicavity \citep{eckart92,lutz93}.  This high ratio indicates that the gas is shocked and the \fethree line emission traces shock destruction of dust \citep{lutz93}. In addition, the high electron temperature, gas density, and unbound gas associated with the minicavity make this source unique in  morphology, kinematics and thermodynamics. These measurements provide compelling evidence that the gas in the bar is tracing an interaction site due to a jet or wind-driven outflow from the direction of Sgr A*. The outflowing material is responsible for sweeping up the interstellar material, creating a cavity within the bar of ionized gas.

On a larger scale, the head-tail sources (e.g., F1 and F3) at position angles (PAs) of $52^\circ$~and $57^\circ$, as well as a striking tower of nonthermal radio emission at PA$\sim50^\circ - 60^\circ$ \citep[see][]{zadeh16} have also been argued to be the sites of an interaction with the atmosphere of dusty stars driven by a collimated outflow by winds or a jet from Sgr A*.  In this picture, the acceleration of the tail to higher positive velocities constrain the geometry of the outflowing material, implying that the cometary sources lie on the far side of Sgr A*. 

The picture we propose is similar to previous collimated outflow models argued in \cite{zadeh12,zadeh16,lutz93}~to explain the origin of highly blue-shifted gas.  The collimated outflow is generated either by the winds of massive young stars \citep[e.g.][]{paumard06, lu09} distributed within $\sim$0.2 pc of Sgr A* or from Sgr A* itself. The  model of a nuclear wind from a cluster of massive stars has to punch through the ionized bar and create the minicavity.  This implies that the cluster wind has to be anisotropic.  However, this scenario appears to be inconsistent with the jet proposed by \cite{li13}~based on a linear X-ray feature that they detected.  The X-ray jet candidate is asymmetric and oriented in the direction towards the Eastern arm, which is roughly perpendicular to the model discussed here.

\begin{figure} 
\includegraphics[width=\textwidth]{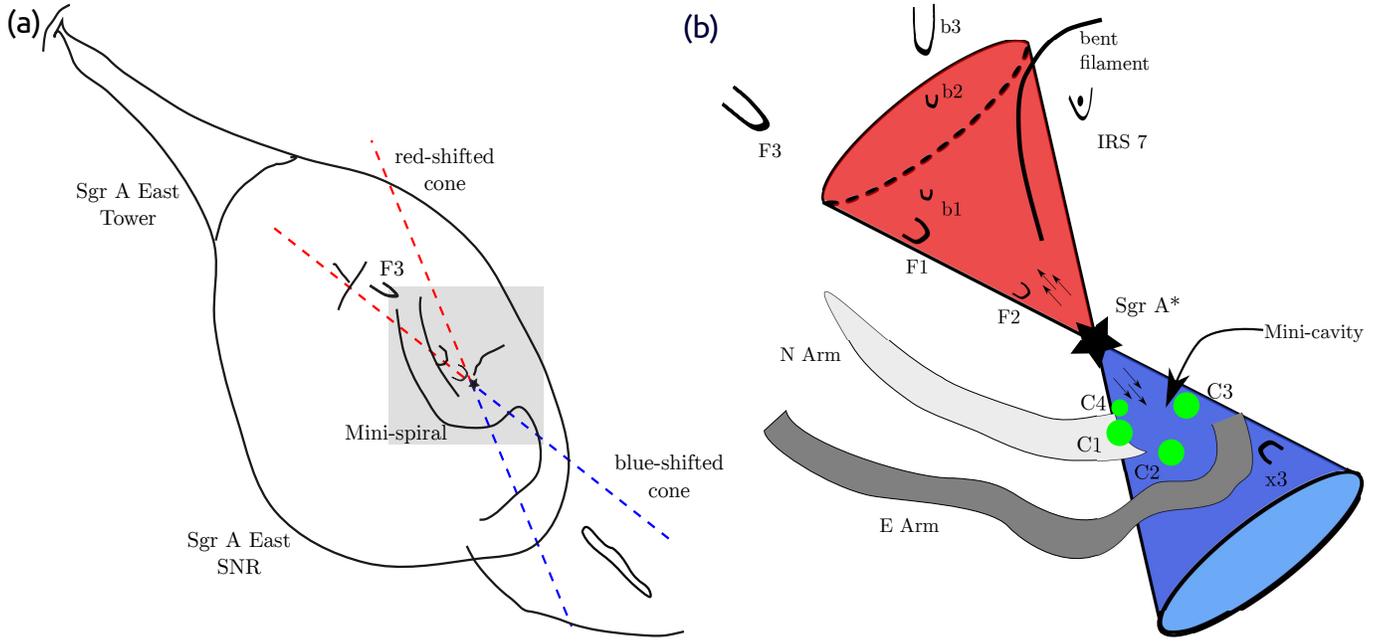}
\caption{\emph{(a)} A broad schematic of the mini-spiral and Sgr A East supernova remnant. \emph{(b)} A schematic showing detail of the shaded region of \emph{a}.  The green/lime circles represent the four C sources discussed in the text.  The red (blue) shading of the cone is the candidate jet-like feature moving away (towards) the observer.  The near side of the jet is blue-shifted and interacts with the mini-cavity.  Meanwhile, the red-shifted component is on the far side of Sgr A* and behind the Northern arm but possibly interacts with the cometary features discussed.}
\label{fig:cartoon}
\end{figure}

If the outflow originates from Sgr A*, the highly disturbed and blue-shifted gas is a result of a mildly relativistic jet symmetrically emanating from Sgr A*. The blue-shifted component of the jet is on the near-side of Sgr A* and punches through the orbiting gas in the ionized bar which results in the observed structure of the minicavity.  This also places the ionized bar on the front side of Sgr A*.  The far-side (red-shifted component) of the jet interacts with the atmosphere of dusty cometary sources toward Sgr A*, creating a more red-shifted tail in the direction of Sgr A*. We also consider the possibility that the far-side of the jet interacts with the Sgr A East supernova remnant and creates the Sgr A East tower, 150$''$ away from Sgr A* with a PA$\sim50^\circ - 60^\circ$ \citep{zadeh16}. In this picture, the Northern arm is aligned close to the plane of Sgr A* so that the interaction of the orbiting ionized gas of the mini-spiral and the jet only occurs on the near-side of Sgr A*. Such an alignment is consistent with previous RL studies using the Keplerian model \citep{zhao09}.  A schematic diagram in Figure~\ref{fig:cartoon}a shows the large-scale structures including the aforementioned Sgr A East tower.  Smaller scale features are displayed in Figure \ref{fig:cartoon}b.  The suggested symmetric jet driven outflow is colored in the schematic to illustrate the radial direction of flow for each side.

\subsection{Mass Outflow Rate}
We now quantify the mass outflow rate required to produce the mini-cavity. The blue-shifted ionized bar lies in the near-side of the plane of the sky before the jet accelerates the gas to more negative velocities.  The change in the velocity of the gas due to this puncture is about 200\kms~over a projected angular distance of $\sim$ 1\arcsec~(see Figure \ref{fig:mcpv}).  If the total mass of the disturbed gas found in Section \ref{sec:density} as $\sim$ 0.052\msol~is utilized, then we can estimate the momentum deposition rate as 0.051\msol~year$^{-1}$\kms.  Adopting a Lorentz factor of $\gamma\sim2.7$ for the mildly relativistic jet, a mass outflow rate of $1.8\times10^{-7}$\msol~year$^{-1}$ would satisfy a jet origin for the mini-cavity.

Alternatively, if the outflow is driven by winds from massive stars, the required mass-loss rate is $4\times10^{-5}$ \msol\, yr$^{-1}$ assuming a wind velocity of $\sim1.25\times10^3$\kms. The global wind accelerating C1 - C4 and creating the mini-cavity has to be anisotropic with a total mass loss rate much higher than estimated here.

The cometary sources can also be utilized to check for consistencies in a collimated outflow model with similar arguments as above.  However, the change in velocity from the head to the tail over the angular distance (Section \ref{sec:comet} and Table \ref{tab:table}) is used to determine the force or momentum deposition rate imparted onto the cometary source that is moving within the candidate outflow.  For the F sources a total mass of $1.6\times10^{-3}$ and $5.4\times10^{-3}$\msol~was found for F1 and F3 respectively.  

Unlike the mini-cavity structure however, the cometary sources only subtend a small fraction of the collimated outflow since they are more than a few arcseconds from Sgr A*.  We adopt an opening angle for this outflow of 30\degr~determined by considering the position angle difference between C1 and C3.  Similar cometary sources (X3 and X7) have been modeled by \cite{muzic10} where a size of a few hundred AU was determined for the bow shock standoff distance.  Thus, a reasonable estimate to the upper limit of the cross sectional size of the mid-IR stellar source is 50 AU.  These parameters yield total mass outflow rates of $3.8\times10^{-6}$ and $4.3\times10^{-5}$\msol~year$^{-1}$ for a collimated jet.  Whereas a wind driven outflow would have to drive $\sim10^{-4}$\msol~year$^{-1}$.

Both b1 and b2 imply similar mass outflow rates as to the cometary sources.  For a collimated jet, a mass outflow rate of $8.5\times10^{-6}$ and $3.8\times10^{-6}$\msol~year$^{-1}$ is required if it is responsible for b1 and b2 respectively.  On the other hand, b3 requires a much larger outflow rate of $7.4\times10^{-5}$\msol~and a highly unlikely wind mass loss rate of 0.02\msol~year$^{-1}$.  The larger estimates required for b3 are most likely due to its particularly elongated morphology.  However, this most likely suggests b3 is not interacting with the proposed collimated outflow and is caused through different mechanisms.

\subsection{Summary}
Multiple highly blue-shifted disturbed features of ionized gas have been detected. The high velocity sources are associated with the mini-cavity within 2$''$ of Sgr A*. In addition we found the kinematics of cometary sources pointing in the direction of Sgr A*. These measurements suggest a collimated outflow from Sgr A* or an outflowing nuclear wind, possibly produced by the young mass-losing stars near Sgr A*.  A wind-driven outflow is unlikely to explain the origin of the mini-cavity and unbound ionized gas unless the outflow is anisotropic with a position angle of $\sim$50$^\circ$ (and $\sim$230$^\circ$) within 14\asec~of Sgr A*. One implication of the outflowing material is the prevention of gaseous material falling into the accretion disk of Sgr A*, thus reducing the accretion rate. The proposed hypothesis for a jet from Sgr A* can be tested observationally in the future by searching for linearly polarized emission associated with Sgr A* at radio and submm wavelengths. In addition, future high resolution proper motion measurements of C1-C4 towards the mini-cavity combined with radial velocity measurements will help determine if they are unbound.  

\acknowledgements
This work is partially supported by the grant AST-0807400 from the NSF. The National Radio Astronomy Observatory is a facility of the National Science Foundation operated under cooperative agreement by Associated Universities, Inc. This Letter makes use of the following ALMA data: ADS/JAO.ALMA\#2015.A.00021.S. ALMA is a partnership of ESO (representing its member states), NSF (USA) and NINS (Japan), together with NRC (Canada) and NSC and ASIAA (Taiwan), in cooperation with the Republic of Chile. The Joint ALMA Observatory is operated by ESO, AUI/NRAO and NAOJ.

\software{CASA (v5.1.1; \cite{mcmullin07})}

\begin{deluxetable}{ccccccccccc} 
\tabletypesize{\scriptsize} 
\tablecolumns{11} 
\tablewidth{0pt} 
\tablecaption{  \label{tab:table}} 
\tablehead{ 
       & \colhead{$\Delta\alpha$, $\Delta\delta$ (J2000)$^\dagger$}    & \colhead{Offset$^\dagger$}  & \colhead{PA$^\dagger$}  & \colhead{Size}   
 & \colhead{S$_\mathrm{L}$}   & \colhead{$\mathrm{v}_\mathrm{r}$} & \colhead{$\Delta \mathrm{v}_\mathrm{r}$}  
 & \colhead{$n_\mathrm{e}^\ddagger$}  & \colhead{Length} & \colhead{$\left|\Delta \rm V\right|$} \\
\colhead{Source} & \colhead{(arcsec)}        & \colhead{(arcsec)}    & \colhead{(deg)} &  \colhead{(mas)} 
& \colhead{(mJy)}            & \colhead{(km s$^{-1}$)}    & \colhead{(km s$^{-1}$)}    
&  \colhead{($10^4$ cm$^{-3}$)} & \colhead{(arcsec)} & \colhead{(km s$^{-1}$)}}
\startdata 
C1   & --0.68, --2.06 & 2.17 & --163 & 320 & 2.06 & --455 & 37.0 & 3.39 - 4.97 & \\
     &              &      &       &        & 2.24 & --419 & 33.1 &            \\
     &              &      &       &        & 2.92 & --372 & 89.0 &            \\    
C2   & --1.44, --3.11 & 3.43 & --155 & 180 & 3.21 & --388 & 41.1 & 2.99 - 4.39 \\
     &              &      &       &       & 1.28 & --329 & 37.9 &           \\ 
C3   & --1.99, --1.96 & 2.80 & --135 & 420$^\star$ & 4.51 & --381 & 35.7 & 1.85 - 2.72 \\
C4   & --0.99, --1.72 & 1.99 & --149 & 390$^\star$ & 2.01 & --366 & 40.0 & 1.36 - 1.99 \\
F1   &  +3.89, +3.07 & 4.95 & 51.7  & 390$^\star$& 6.42 & 181   & 26.4 & 1.97 - 2.89 & 0.5 & 8\\
%F2   & 40.14 & 27.11 &  &  &    &  & & 54.8 & & \\
F3   &   +10.7, +7.08  & 12.8 & 56.5  & 580  & 11.8 & 196   & 22.1 & 2.01 - 2.94 & 0.7 & 6.7 \\
b1   & +4.29, +4.42 & 6.16 &   35.7  & 480  & 2.92 & 166   & 31.0 & 1.30 - 1.91 & 0.3& 6.6 \\
b2   & +5.04, +8.20 & 9.63 & 31.6  & 530  & 7.09 & 152   & 25.6 &  1.75    - 2.57  & 0.69 & 3.67  \\
b3   & +6.76, +11.7 & 13.5 & 30.0   & 930  & 10.4 & 164 & 22.9 &  1.51- 2.22 & 2.1 & 8.2 \\
\enddata
\vspace{-0.5mm}
\tablecomments{$\dagger$ With respect to Sgr A* (\radec{40}{038}{28}{069}).\\
$\ddagger$ Assuming electron temperatures ranging from 7.5$\times10^3$ K - 12.5$\times10^3$ K.\\
$^\star$ Unresolved, given size is an upper limit convolved with the beam.}
\end{deluxetable}

\clearpage
\bibliographystyle{apj}
\bibliography{unboundrrl}

\end{document}